\documentclass[letterpaper]{article}
\usepackage{aaai16}
\usepackage{times}
\usepackage{helvet}
\usepackage{courier}
\usepackage{amsthm,lipsum}
\usepackage{mathrsfs}
\usepackage{graphicx}
\usepackage{amsmath}
\usepackage{algorithm}
\usepackage{algorithmic}
\usepackage{multirow}
\usepackage{afterpage}
\title{Lift-Based Bidding in Ad Selection}
\author{Jian Xu$^*$, Xuhui Shao, Jianjie Ma, Kuang-chih Lee, Hang Qi, Quan Lu\\ $^*$ TouchPal Inc., 1172 Castro St, Mountain View, CA 94040 \\ Yahoo Inc., 701 First Ave, Sunnyvale, CA 94089 \\ jian.xu@cootek.cn, \{xshao,jianma,kclee,hangqi,qlu\}@yahoo-inc.com}

\newtheoremstyle{slplain}
  {.5\baselineskip\@plus.2\baselineskip\@minus.2\baselineskip}
  {.5\baselineskip\@plus.2\baselineskip\@minus.2\baselineskip}
  {\slshape}
  {10pt}
  {}
  {.}
  { }
  {}
\theoremstyle{slplain}
\newtheorem{definition}{\textsc{Definition}}
\newtheorem{theorem}{\textsc{Theorem}}
\newtheorem{lemma}{\textsc{Lemma}}

\newtheoremstyle{slremark}
  {.5\baselineskip\@plus.2\baselineskip\@minus.2\baselineskip}
  {.5\baselineskip\@plus.2\baselineskip\@minus.2\baselineskip}
  {}
  {10pt}
  {}
  {.}
  { }
  {}
\theoremstyle{slremark}
\newtheorem{example}{\textsc{Example}}
\renewcommand{\qed}{\hfill\rule{1ex}{1ex}}

\begin{document}
\maketitle

\begin{abstract}
\begin{quote}
Real-time bidding has become one of the largest online advertising markets in the world. Today the bid price per ad impression is typically decided by the expected value of how it can lead to a desired action event to the advertiser. However, this industry standard approach to decide the bid price does not consider the actual effect of the ad shown to the user, which should be measured based on the performance lift among users who have been or have not been exposed to a certain treatment of ads. In this paper, we propose a new bidding strategy and prove that if the bid price is decided based on the performance lift rather than absolute performance value, advertisers can actually gain more action events. We describe the modeling methodology to predict the performance lift and demonstrate the actual performance gain through blind A/B test with real ad campaigns. We also show that to move the demand-side platforms to bid based on performance lift, they should be rewarded based on the relative performance lift they contribute.
\end{quote}
\end{abstract}

\section{Introduction}
\label{sec:intro}
Online advertising is one of the fastest growing industries with \$58 billion total spend projected in 2015 in US alone. One of the most significant trends in online advertising in recent years is real-time bidding (RTB), or sometimes more broadly referred to as programmatic buying. In RTB, advertisers have the ability of making decisions whether and how much to bid for every impression that would lead to the best expected outcome. It is analogous to stock exchanges in that data-driven algorithms are used to automatically buy and sell ads in real-time. The bidding algorithm can use the contextual and user behavioral data to select the best ads in order to optimize the effectiveness of online advertising.\\
\indent Demand-Side Platforms (DSPs) are thus created to help advertisers manage their campaigns and optimize their real-time bidding activities. DSPs offer different pricing models, such as cost per impression (CPM\footnote{Sometimes refers to cost per thousand impressions to make the numbers easier to manage.}), cost per click (CPC), and cost per action (CPA). This paper focuses on the CPA pricing model as it is the most challenging problem. State-of-the-art DSPs that support such CPA pricing model typically convert an advertiser's CPA bid to an expected cost per impression (eCPM) bid in order to participate in the RTB auctions where the winning ad is chosen based on the highest bid \cite{mcafee2011design}. In pure second price auctions, theoretically the optimal bidding strategy is truth-telling. Therefore, the prevalent practice to derive such eCPM bid is estimating the Action Rate (AR) which is the probability that the impression will lead to a desired action and multiplying it by the CPA bid (i.e., eCPM=AR$\times$CPA).\\
\indent However, such bidding strategy neglects the probability that a user will take the desired action even if the impression is not shown. For example, a loyal Pampers customer will make further purchases even if not exposed to any Pampers ad. An analogy can be found in the political marketing. As early as in 2012, Obama's presidential campaign already focused on the swing voters by quantifying how easily they can be shifted to vote for the Democrat \cite{rutenberg2013data}. It is surprising that the online advertising market has lagged behind at this point. We argue that the prevalent bidding strategy is categorically suboptimal by design.

\subsection{Motivating Examples}
The above bidding strategy calculates the bid price based on the AR (proxy for \emph{value} to the advertiser) of a user. The problem is that campaign budget would be spent on users who already have high ARs instead of those who could have been greatly influenced by the ads (i.e., those with high AR \emph{lift} because of the ads). The difference is best illustrated by the examples below.
\begin{example}[\textsc{Value-based Bidding}]
\label{ex:value-based}
Suppose a DSP is bidding on behalf of an advertiser to acquire impressions and the advertiser's CPA=\$100. Suppose there are two ad requests from user $a$ and $b$ respectively. The AR of user $a$ is 0.04 if she is shown the advertiser's ad, otherwise the AR is 0.03. User $b$ has an AR of 0.02 if she is shown the ad, otherwise the AR is 0.001. \\
\indent Based on the common practice in the industry, the bid prices should be the absolute ARs assuming ads are shown times CPA. So the the bid prices for $a$ and $b$ are $0.04\times\$100=\$4$ and $0.02\times\$100=\$2$ respectively. Suppose the highest bid prices from other competitors are equally $\$3.5$ for both $a$ and $b$, which means the advertiser will win the auction for $a$ while lose the auction for $b$. In this case, the expected total number of actions the advertiser can have is $0.04+0.001=0.041$, and the inventory cost of the DSP is $\$3.5$. Since advertiser only pays for impressions that lead to actions, the expected revenue of the DSP is $\$4$. \qed
\end{example}
\begin{example}[\textsc{Lift-based Bidding}]
\label{ex:lift-based}
Let us continue with Example \ref{ex:value-based}. It is not difficult to see that user $b$ should be more preferable to the advertiser because the advertiser could expect a more significant AR lift from her (the AR lift of $a$ is $0.04-0.03=0.01$ while $b$ is $0.02-0.001=0.019$). \\
\indent If the bid prices the DSP places on $a$ and $b$ are proportional to the AR lifts, for instance $\$2$ and $\$3.8$, then the advertiser will win the auction for $b$ instead of $a$. In this case, the expected total number of actions the advertiser can have is $0.03+0.02=0.05$ - better than that in Example \ref{ex:value-based}. The inventory cost of the DSP is also $\$3.5$, but the expected revenue of the DSP becomes $\$2$ because the advertiser only pays for impressions that lead to actions. It results in a negative profit to the DSP. \qed
\end{example}

The above toy examples show that the advertiser eventually gets more actions when bid based on AR lift than based on absolute AR. In addition, they also show that the advertiser's marketing objective to maximize actions is not aligned with the DSP's interest. 

\subsection{Our Contribution}
In this paper, we advocate for an industry-wide transition from \emph{value-based bidding} to \emph{lift-based bidding}. That is, the bid price should be based on the AR lift instead of the absolute AR. Our contributions can be summarized as follows: 
\begin{itemize}
\item We propose the concept of lift-based bidding, which we prove both mathematically and empirically to be a better bidding strategy than value-based bidding in terms of maximizing advertiser benefits.
\item We describe a simple yet effective modeling methodology to predict AR lift. Online A/B test with real ad campaigns backed up our concepts and techniques.
\item We point out that the advertiser's marketing objective is not aligned with the DSP's interest. To move the DSPs to lift-based bidding, they should be rewarded based on the relative performance lift they contribute.
\end{itemize} 

\section{Value-Based Bidding vs. Lift-Based Bidding}
In this section, we prove that lift-based bidding is a better strategy than value-based bidding in terms of maximizing advertiser benefits. However, it would be opposed by DSPs under the industry-standard last-touch attribution model.

\begin{definition}[\textsc{AR, background AR, and AR Lift}]
Given an ad request $q$ from a user $u$ and an advertiser $A$, we define AR w.r.t. $(q,\ u,\ A)$ as the probability that $u$ will take the desired action defined by $A$ after the ad of $A$ is served to $q$, background AR w.r.t. $(q,\ u,\ A)$ as the probability that $u$ will take the desired action if the ad of $A$ is not served to $q$, and AR lift as the difference between AR and background AR. We denote by $p$ the AR, $\Delta p$ the AR lift, and $p-\Delta p$ the background AR if no further specification is made. \qed
\end{definition}
The common practice in the industry is to bid CPA$\times p$. We generalize this practice and define value-based bidding as follows:
\begin{definition}[\textsc{Value-Based Bidding}]
Let $p$ be the AR of a user if the advertiser's ad is shown, value-based bidding places a bid price of $\alpha\times p$ to acquire an impression from this user for the advertiser, where $\alpha > 0$. \qed
\end{definition}
However, examples in the \emph{Introduction} section show that such bidding strategy does not necessarily optimize the overall campaign performance for the advertiser. As we advocate for focusing on the AR lift instead of absolute AR, we propose the concept of lift-based bidding in which the bid price is proportional to the AR lift.
\begin{definition}[\textsc{Lift-Based Bidding}]
Let $\Delta p$ be the AR lift of a user if the advertiser's ad is shown, lift-based bidding places a bid price of $\beta\times \Delta p$ to acquire an impression from this user for the advertiser, where $\beta > 0$. \qed
\end{definition}
In the CPA pricing model, a DSP is rewarded based on the number of actions attributed to it. Advertisers finally pay DSPs based on the amount of actions attributed to them. The industry standard attribution model is \emph{last-touch attribution}.
\begin{definition}[\textsc{Last-Touch Attribution}]
An advertiser attributes the full credit of an observed user action to the DSP which delivered the last relevant ad impression to the user. \qed
\end{definition}
Suppose there are two DSPs $DSP_1$ and $DSP_2$ bidding on behalf of the same advertiser at the same time. $DSP_1$ practices value-based bidding while $DSP_2$ executes lift-based bidding. Let $u_i$ be the user of the $i$-th ad request, $p_i$ be the AR if the advertiser's ad is shown to $u_i$, and $\Delta p_i$ be the AR lift because of the ad impression. To simplify the discussion, let us assume every ad request is from a different user and there are no additional candidates in the auctions. In an ad exchange marketplace with pure second-price auction, we have
\begin{lemma}
\label{lemma}
$DSP_1$ wins the auction for $u_i$ at the cost of $\beta\times \Delta p_i$ if $\alpha \times p_i > \beta \times \Delta p_i$; $DSP_2$ wins the auction for $u_i$ at the cost of $\alpha \times p_i$ if $\alpha \times p_i < \beta \times \Delta p_i$.\qed
\end{lemma}

\begin{theorem}
With the last-touch attribution model, $DSP_2$ yields more actions than $DSP_1$ for the advertiser when the advertiser attributes the same amount of actions to them \footnote{The condition of $DSP_1$ and $DSP_2$ getting equal attributions from the advertiser is an important setup to illustrate that $DSP_2$ will in fact produce relatively more actions for the advertiser. This setup condition can be achieved by adjusting the winning landscape through the parameters of $\alpha$ and $\beta$}.
\begin{proof}
Let $i$ be the index of all the users, $j$ be the index of those users that $DSP_1$ wins (i.e., $\alpha \times p_j > \beta \times \Delta p_j$), and $k$ be the index of those users that $DSP_2$ wins (i.e., $\alpha \times p_k < \beta \times \Delta p_k$). It is straightforward to see that the expected number of actions to be attributed to $DSP_1$ and $DSP_2$ are $\sum_j p_j$ and $\sum_k p_k$ respectively. The expected number of actions if only $DSP_1$ is considered can be decomposed as two parts: sum of the ARs of users that $DSP_1$ wins, and sum of the background ARs of users that $DSP_1$ loses. So it becomes $\sum_j p_j + \sum_k (p_k-\Delta p_k)$. Similarly, the expected number of actions if only $DSP_2$ is considered is $\sum_j (p_j - \Delta p_j) + \sum_k p_k$.\\
\indent Therefore, let $\mathscr{A}_1$ ($\mathscr{A}_2$) be the expected number of actions per attributed action if only $DSP_1$ ($DSP_2$) is considered, we have
\begin{equation}
\mathscr{A}_1=\frac{\sum_j p_j + \sum_k (p_k-\Delta p_k)}{\sum_j p_j}
\end{equation}
\begin{equation}
\mathscr{A}_2=\frac{\sum_j (p_j - \Delta p_j) + \sum_k p_k}{\sum_k p_k}
\end{equation}

When the same amount of actions is attributed to $DSP_1$ and $DSP_2$ (i.e., $\sum_j p_j=\sum_k p_k$), by swapping the denominators and consolidating the numerators in Equation 1 and 2, noticing $\alpha \times p_j > \beta \times \Delta p_j$ and $\alpha \times p_k < \beta \times \Delta p_k$, we have
\begin{equation}
\mathscr{A}_1=\frac{\sum_i p_i - \sum_k \Delta p_k}{\sum_k p_k} < \frac{\sum_i p_i}{\sum_k p_k} - \frac{\alpha}{\beta}
\end{equation}
\begin{equation}
\mathscr{A}_2=\frac{\sum_i p_i - \sum_j \Delta p_j}{\sum_j p_j} > \frac{\sum_i p_i}{\sum_j p_j} - \frac{\alpha}{\beta}
\end{equation}

Since $\sum_j p_j=\sum_k p_k$, it is obvious that $\mathscr{A}_1 < \mathscr{A}_2$. 
\end{proof}
\end{theorem}

\begin{theorem}
With the last-touch attribution model, $DSP_2$ costs more than $DSP_1$ when the advertiser attributes the same amount of actions to them. 
\begin{proof}
Again, let $i$ be the index of all the users, $j$ be the user index such that $\alpha \times p_j > \beta \times \Delta p_j$, and $k$ be the user index such that $\alpha \times p_k < \beta \times \Delta p_k$. The expected number of actions to be attributed to $DSP_1$ and $DSP_2$ are $\sum_j p_j$ and $\sum_k p_k$ respectively. The cost of $DSP_1$ and $DSP_2$ are $\sum_j \beta \times \Delta p_j$ and $\sum_k \alpha \times p_k$ respectively. Therefore let $\mathscr{C}_1$ and $\mathscr{C}_2$ be the cost per attributed action of $DSP_1$ and $DSP_2$ respectively,  we have
\begin{equation}
\mathscr{C}_1=\frac{\sum_j \beta \times \Delta p_j}{\sum_j p_j}
\end{equation}
\begin{equation}
\mathscr{C}_2=\frac{\sum_k \alpha \times p_k}{\sum_k p_k} 
\end{equation}

Noticing $\alpha \times p_j > \beta \times \Delta p_j$, we have $\mathscr{C}_1 < \beta \times \frac{\alpha}{\beta} = \alpha$. It is apparent that $\mathscr{C}_2=\alpha$. Therefore $\mathscr{C}_1 < \mathscr{C}_2$. 
\end{proof}
\end{theorem}
We have shown that lift-based bidding benefits the advertisers but the higher cost per attributed action may undermine the interests of the DSPs. The root cause of this conflict is that the industry-standard attribution model does not attribute actions fairly to the DSPs. Researchers have pointed out that an action should be attributed to multiple touch points in a data driven fashion \cite{shao2011data} or based on causal lift \cite{dalessandro2012causally}. In the \emph{Attribution and Bidding} section, we discuss the relationship between attribution models and bidding strategies. We show that to move the DSPs from value-based bidding to lift-based bidding, they should be rewarded based on the relative action lift they contributed to the final actions. 

\section{Lift-Based Bidding in Action}
\subsection{Predicting AR Lift}
To implement lift-based bidding, it is important to estimate the AR lift. One may think of building a machine learning model to predict the lift directly. However, since the real ad serving logs contain only instances that an ad is either shown or not shown, it is theoretically impossible to have the true AR lift data for modeling. Therefore, in order to predict the AR lift, we strive to estimate both the ARs assuming the ad is shown or not shown respectively. \\
\indent Formally, let $a$ be an ad, $s$ be the state of a user at ad request time, and $s_+(a)$ be the state of the user if $a$ is shown. Conceptually, $s$ consists of the user's demographic status, timestamped past events including page views, searches, ad views/clicks, and anything that describes the user state at ad request time. The only difference between $s$ and $s_+(a)$ is the ad impression of $a$. Let $p(action|s)$ be the AR of the user if $a$ is not shown and $p(action|s_+(a))$ be the AR if $a$ is shown, the AR lift is 
\begin{equation}
\Delta p = p(action|s_+(a)) - p(action|s)
\end{equation}
\indent Considering a specific ad request instance in the ad serving log, the ad was either shown or not shown. Therefore, either case will be absent in the modeling data. We address this challenge by establishing a model that has sufficient generalization capability. More specifically, we use a function $F$ to map a state to a set of features shared among different instances. Then a single and generic AR prediction model $\hat{P}$ is built upon the derived feature set and the AR lift can be estimated as
\begin{equation}
\widehat{\Delta p} = \hat{P}(action|F(s_+(a))) - \hat{P}(action|F(s))
\end{equation}
\indent The difference between $F(s_+(a))$ and $F(s)$ is reflected by different feature values induced from $a$. At ad serving time when $\Delta p$ is to be estimated, if for instance we consider impression frequency of $a$ as a feature, the feature value in $F(s_+(a))$ should be greater than that in $F(s)$ by one.
\subsection{Model Training}
Our task is to train a generic AR prediction model $\hat{P}$ to give AR estimations for both cases when an ad is shown or not shown. Existing AR prediction models in the literature are trained based on the post-view \cite{KLee:EstimateCVRTurn} or post-click \cite{rosales2012post} actions. That is, the training samples are collected from only those impression events or click events. For example, in post-view action modeling, each impression event will trigger a training sample which is labeled as positive if there is an action followed. However, such methodology is not preferred in our scenario for several reasons. \\
\indent First, since the training samples from only impression or click events are not representative to those cases when the ad is not shown, models trained upon these samples are not generalized enough to predict $p(action|s)$. Second, even for predicting  $p(action|s_+(a))$, leveraging training samples from only impression or click events still suffers from survival bias. In the RTB marketplace, impressions are purchased through public auctions. Therefore, impressions and clicks are available from only those winning auctions. Such survival bias is prevalent in click modeling. In action modeling, it could be avoided because actions can happen even there was no impression showed. Third, our observation from real ad campaigns shows that for the majority (usually $>$90\%) of actions, we have not shown any ad of the advertiser before\footnote{Advertisers usually report actions via action pixels placed on their websites or apps. Therefore DSPs can have the full action set.}. In other words, if training samples are only generated from impression or click events, the majority of actions (positive samples) are not leveraged for modeling.\\
\begin{figure}
\centering
\includegraphics[width=8cm]{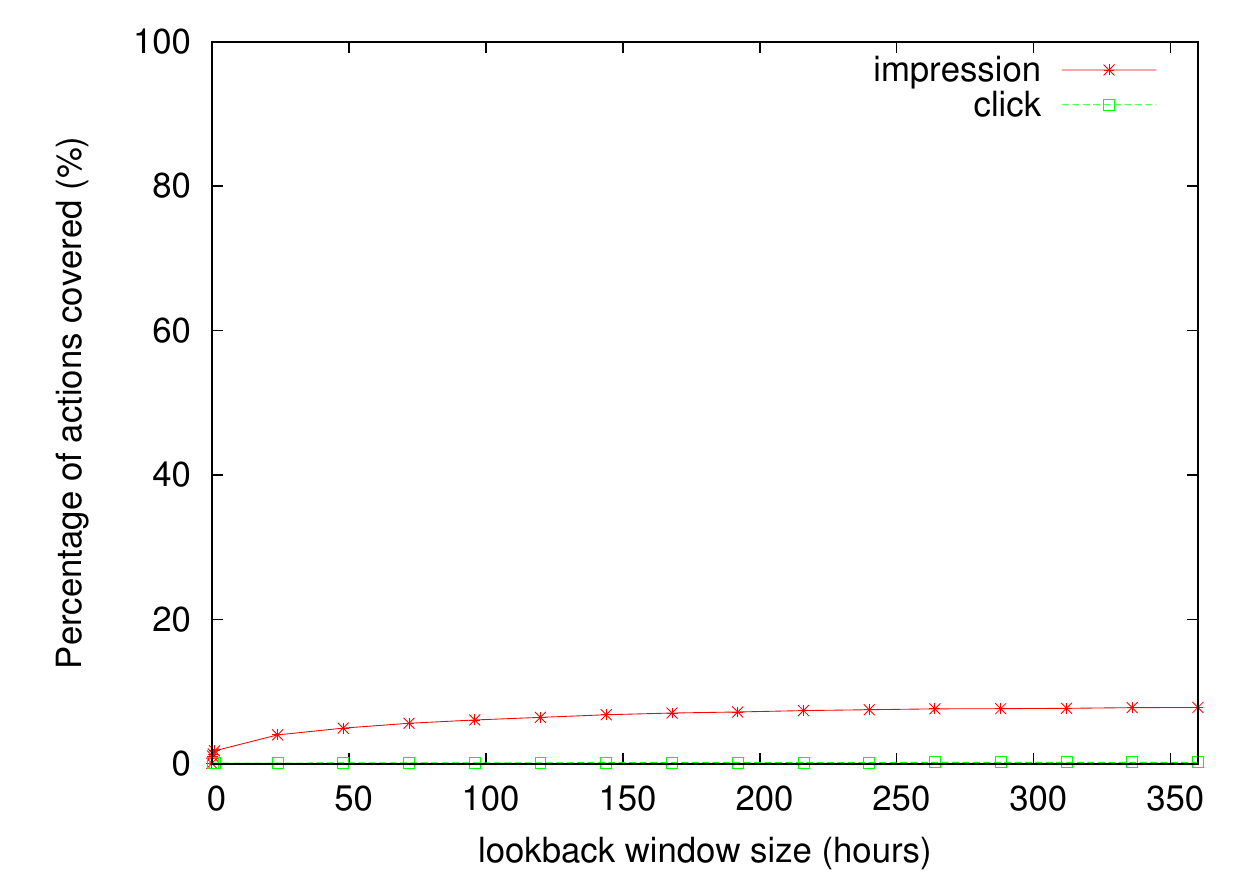}
\caption{Less than 10\% of the reported actions have precedent ad impressions from the same advertiser within the lookback window. Training samples generated from only impression and/or click events miss a large portion of the informative actions.}
\label{fig:coverage}
\end{figure}
\indent Therefore, we take a different approach that we train the AR prediction model upon the whole population. Training samples are generated from every user's timeline instead of from merely impression/click events. To mimic the true action distribution, we first randomly select a user $u$ weighted by its ad request frequency. Then a random timestamp $ts$ is chosen on $u$'s timeline and a training sample is generated based on $u$ and $ts$. If $u$ has at least one action in the \emph{action window} (denoted by $(ts, ts+aw]$), the sample is labeled as positive and otherwise negative, where $aw$ is the action window size as long as several hours to several days based on business definitions. Then features are generated within the \emph{feature window} (denoted by $(ts - fw, ts]$), where $fw$ is the feature window size. \\
\indent Raw input variables for feature generation include user historical profile within the feature window with such as page views, ad impressions, clicks, searches and mobile app based events. Each user event is used as a point-in-time to generate features. Table \ref{tab:features} is a list of different types of features we generated for AR prediction. At serving time, ad request details such as geo-location, web page or mobile app being visited are folded in these features so that the run-time context can also be leveraged for prediction. For example, if the recency of visiting Yahoo! homepage is a feature, an ad request from Yahoo! homepage will set this feature value as ``most recent". \\ 
\indent The sample generation terminates when all the action events have been involved or the positive samples are sufficient. Once the training samples are gathered, we train a Gradient-Boosting-Decision-Tree (GBDT) model to predict the rank order and then calibrate using isotonic regression to translate a GBDT score to an AR. Please note that we utilize our in-house GBDT tool with distributed training capability for modeling; however, other proper machine learning models can also be applied. \\
\begin{table}[t]
\scriptsize
\centering
\begin{tabular}{|c|p{80pt}|p{80pt}|}
\hline
{\bf Source} & {\bf Feature name}                                                                                                                                                                                            & {\bf Note}                                                                                                                                                                  \\ \hline
Behaviors            & IMP\_FREQ\_ADV IMP\_RNCY\_ADV CLK\_FREQ\_ADV CLK\_RNCY\_ADV PV\_FREQ\_TOPIC PV\_RNCY\_TOPIC SRCH\_FREQ\_TOPIC SRCH\_RNCY\_TOPIC & Impression and click frequency/recency from each advertiser. Page-view and search frequency/recency of each topic (pages and search queries are segmented into several semantic topics.) \\ \hline
Demographics         & AGE\_GROUP\newline GENDER\newline GEO\_AREA  & Ages are mapped into several age groups. User's geographic location is at some moderate resolution level.                                                                           \\ \hline
Mobile               & INST\_FREQ\_APP INST\_RNCY\_APP USE\_FREQ\_APP USE\_RNCY\_APP  & Installation and usage frequency/recency of each mobile app.                                                                                                                       \\ \hline
\end{tabular}
\caption{Features generated for AR modeling.}
\label{tab:features}
\end{table}
\subsection{Fitting Lift-Based Bidding in the Market}
\label{sec:fitting}
Conventional value-based bidding calculates the bid price by multiplying predicted absolute AR by advertiser CPA. In lift-based bidding, it is not proper to simply multiply AR lift by the same advertiser CPA. Otherwise one can seldom win any auction if the majority of the other competitor DSPs are still practicing value-based bidding. Recall that in lift-based bidding the bid price is proportional to the AR lift i.e., $\beta \times \Delta p$. Our selection of $\beta$ is 
\begin{equation}
\beta = \frac{\overline{p}}{\overline{\Delta p}} \times \text{CPA}
\end{equation}
where $\overline{p}$ is the population mean of AR and $\overline{\Delta p}$ is the population mean of AR lift. The idea is straightforward: if the advertiser is willing to pay CPA for each action in the conventional way, then each incremental action should be paid at the price of $\frac{\overline{p}}{\overline{\Delta p}} \times$CPA if only incremental actions need to get paid.

\section{Blind A/B Test with a Real DSP}
To empirically prove our proposed concepts, we set up A/B test experiments on Yahoo's Demand-Side Platform. We first randomly split users into three equal-sized groups. Then we created three bidders: a \emph{passive bidder} which always places a zero bid, a value-based bidder, and a lift-based bidder. We selected five advertisers to participate in the test. To be fair, each advertiser's budget was evenly split and assigned to the value-based bidder and lift-based bidder respectively (the passive bidder would not spend any budget). Each advertiser's campaign ran for one week and their budgets were all spent out which means value-based bidder and lift-based bidder spent the same amount of budget. We counted the number of actions\footnote{Action pixel fires to be more accurate. Even in the passive group, there can still be actions reported via action pixel fires on advertiser's website or app.} observed in each group in a three weeks window from the campaign start date. The results shown in Table \ref{tab:ab1}, \ref{tab:ab2}, \ref{tab:ab3} and \ref{tab:ab4} backed up our claims and methods.

\afterpage{
\begin{table*}[t!]
\center
\scalebox{0.85}{
\begin{tabular}{|c|c|c|c|c|c|c|}
\hline
\multirow{2}{*}{Adv} & \multicolumn{2}{c|}{passive bidder}      & \multicolumn{2}{c|}{Value-based bidder} & \multirow{2}{*}{Incremental action} & \multirow{2}{*}{\textbf{Action lift}} \\ \cline{2-5}
       & \texttt{\texttt{\#}} imps & \texttt{\texttt{\#}} actions & \texttt{\texttt{\#}} imps           & \texttt{\texttt{\#}} actions   &                 &               \\ \hline
1     & 0           & 642          & 53,396            & 714            & 72            & \textbf{11.2\%}   \\ 
2     & 0           & 823          & 298,333          & 896            & 73            & \textbf{8.9\%}     \\ 
3     & 0           & 1,438       & 11,048,583     & 1,477         & 39            & \textbf{2.7\%}     \\ 
4     & 0           & 1892        & 3,915,792       & 2,016         & 124          & \textbf{6.6\%}     \\ 
5     & 0           & 5,610       & 6,015,322     & 6,708         & 1,098       & \textbf{19.6\%}    \\  
\hline
\end{tabular}
}
\caption{Blind A/B test on five pilot advertisers - Value-based bidder v.s. Passive bidder. }
\label{tab:ab1}
\end{table*}

\begin{table*}[t!]
\center
\scalebox{0.85}{
\begin{tabular}{|c|c|c|c|c|c|c|}
\hline
\multirow{2}{*}{Adv} & \multicolumn{2}{c|}{passive bidder}      & \multicolumn{2}{c|}{Lift-based bidder} & \multirow{2}{*}{Incremental action} & \multirow{2}{*}{\textbf{Action lift}} \\ \cline{2-5}
                                        & \texttt{\#} imps & \texttt{\#} actions & \texttt{\#} imps   & \texttt{\#} actions  &                                                 &                                       \\ \hline
1                                       & 0                       & 642                 & 59,703                    & 826                  & 184                                             & \textbf{28.7\%}                                \\ 
2                                       & 0                       & 823                 & 431,637                   & 980                  & 157                                             & \textbf{19.1\%}                                \\ 
3                                       & 0                       & 1,438               & 11,483,360                & 1509                 & 71                                              & \textbf{4.9\%}                                 \\ 
4                                       & 0                       & 1892                & 4,368,441                 & 2,471                & 579                                             & \textbf{30.6\%}                                 \\ 
5                                       & 0                       & 5,610               & 8,770,935                & 8,291                 & 2,681                                              & \textbf{47.8\%}                                 \\  
\hline
\end{tabular}
}
\caption{Blind A/B test on five pilot advertisers - Lift-based bidder v.s. Passive bidder. }
\label{tab:ab2}
\end{table*}

\begin{table*}[t!]
\center
\scalebox{0.84}{
\begin{tabular}{|c|c|c|c|c|c|c|c|c|c|}
\hline
\multirow{2}{*}{Adv} & \multicolumn{3}{c|}{Value-based bidder}   & \multicolumn{3}{c|}{Lift-based bidder}   & \multirow{2}{*}{\begin{tabular}[c]{@{}c@{}}Action lift\end{tabular}} & \multirow{2}{*}{\textbf{\begin{tabular}[c]{@{}c@{}}Lift-over-lift\end{tabular}}} \\ \cline{2-7}
                     & \texttt{\#} imps    & \texttt{\#} actions & \begin{tabular}[c]{@{}c@{}}Action lift \\ (vs. passive)\end{tabular} & \texttt{\#} imps    & \texttt{\#} actions & \begin{tabular}[c]{@{}c@{}}Action lift\\ (vs. passive)\end{tabular} &                                                                         &                                                                                                                                                                        \\ \hline
1                    & 53,396     & 714        & 11.2\%                                                                    & 59,703     & 826        & 28.7\%                                                                  & 13.6\%                                                                  & \textbf{156\%}                                                                                                                                        \\ 
2                    & 298,333    & 896        & 8.9\%                                                                     & 431,637    & 980        & 19.1\%                                                                  & 9.4\%                                                                   & \textbf{115\%}                                                                                                                                          \\ 
3                    & 11,048,583 & 1,477      & 2.7\%                                                                     & 11,483,360 & 1509       & 4.9\%                                                                   & 2.2\%                                                                   & \textbf{82\%}                                                                                                                                           \\ 
4                    & 3,915,792  & 2,016      & 6.6\%                                                                     & 4,368,441  & 2,471      & 30.6\%                                                                  & 22.6\%                                                                  & \textbf{367\%}                                                                                                                                   \\ 
5                    & 6,015,322  & 6,708      & 19.6\%                                                                     & 8,770,935  & 8,291      & 47.8\%                                                                  & 23.6\%                                                                  & \textbf{144\%}                                                                                                                                   \\
\hline
\end{tabular}
}
\caption{Lift-based bidding vs. Value-based bidding - Advertiser's perspective. ``Action lift" is the absolute \texttt{\#} actions difference between lift-based bidder and value-based bidder. ``Lift-over-lift" is comparing the their action lifts over passive bidder.}
\label{tab:ab3}
\end{table*}

\begin{table*}[t!]
\center
\scalebox{0.88}{
\begin{tabular}{|c|c|c|c|c|c|c|c|c|c|}
\hline
\multirow{2}{*}{Adv} & \multicolumn{3}{c|}{Value-based bidder}                                                            & \multicolumn{3}{c|}{Lift-based bidder}                                                           & \multirow{2}{*}{\begin{tabular}[c]{@{}c@{}} \textbf{Inventory-}\\\textbf{cost diff}\end{tabular}} & \multirow{2}{*}{\begin{tabular}[c]{@{}c@{}}\textbf{Cost-per-}\\\textbf{imp diff}\end{tabular}} \\ \cline{2-7}
                     & \texttt{\#} imps    & \texttt{\#} attrs & Inventory cost & \texttt{\#} imps    & \texttt{\#} attrs & Inventory cost &                                                                         &       \\ \hline
1                    & 53,396     & 50        & \$278.73       & 59,703     & 50        & \$300.31       & \textbf{7.7\%}     & \textbf{-3.6\%}           \\ 
2                    & 298,333   & 80        & \$1,065.05    & 431,637   & 80        & \$1,467.57    & \textbf{37.8\%}        & \textbf{-4.8\%}           \\ 
3                    & 11,048,583 & 240   & \$25,522.22  & 11,483,360 & 240    & \$25,837.56  & \textbf{1.2\%}       & \textbf{-2.6\%}             \\ 
4                    & 3,915,792  & 200    & \$10,846.74  & 4,368,441  & 200     & \$11,183.21  & \textbf{3.1\%}    & \textbf{-7.6\%}          \\ 
5                    & 6,015,322  & 500    & \$19,296.51  & 8,770,935  & 500     & \$23,501.90  & \textbf{21.8\%}    & \textbf{-16.5\%}          \\ 
\hline
\end{tabular}
}
\caption{Lift-based bidding vs. Value-based bidding - DSP's perspective. Both bidders spent out equal amount of assigned budget, so the \texttt{\#} attributions are always the same. Cost-per-impression is the inventory cost averaged by \texttt{\#} impressions.}
\label{tab:ab4}
\end{table*}
}

\indent First, by comparing passive bidder and value-based bidder (Table \ref{tab:ab1}), it is easy to demonstrate that a user may take actions even if no impression is shown. Although showing ads to users did help lift the action yield, the number of actions when no ad is shown is already significant. This is not surprising because advertisers typically run campaigns through multiple channels such as TV, magazine, internet, etc simultaneously. Even if a user is not shown any display ad by our DSP, she can still be influenced by other touch points. This is exactly why lift-based bidding is more preferable since it tries to maximize the effectiveness of display ads by taking background AR into account. Table \ref{tab:ab2} shows the comparison between passive bidder and lift-based bidder. \\
\indent Second, from advertiser's perspective, lift-based bidder generated more actions than value-based bidder with the same amount of budget. This result is observed consistently among all the five advertisers (Table \ref{tab:ab3}). Since the background actions prevalently exist, a fairer comparison should be comparing their action lifts over background actions, which we call \emph{lift-over-lift}. Take Advertiser 1 for example, value-based bidder generated 11.2\% more actions than the background actions while lift-based bidder yielded 28.7\% more. In this case the lift-over-lift measure is $(28.7-11.2)/11.2=156\%$. Lift-based bidding dramatically increases the \emph{incremental} actions compared to value-based bidding so the lift-over-lift measure is very significant.\\
\indent Third, from DSP's perspective, lift-based bidder resulted in more inventory cost than value-based bidder when the same number of actions were attributed to them. Recall that advertiser only pays for each attributed action at the price of CPA and the two bidders spent the same amount of budget. So the number of attributions are the same. From Table \ref{tab:ab4} we observe the inventory cost of lift-based bidder is consistently higher than that of value-based bidder. \\
\indent Lastly and interestingly, we have observed increased number of impressions when comparing lift-based bidding to value-based bidding. Even though the overall inventory cost is higher, the effective cost per impression is lower for lift-based bidder than value-based bidder. The lift-based bidder does not always compete with other bidders for those high AR users. Instead, it tries to acquire users who are more likely to be influenced. Therefore it has the advantage of avoiding competition and acquiring more impressions at a lower cost per impression. \\
\indent The above results backed up our concepts and techniques. Since the lift-based bidder took the risk of higher cost while as we have seen it actually benefited the advertisers, advertisers should think of a more reasonable attribution model to align the DSPs' benefits with their marketing objectives.

\section{Attribution and Bidding}
\label{sec:attribution}
We have proved that DSPs may be opposed to lift-based bidding because of higher cost per attributed action. The root cause is that they are not rewarded based on the action lift they contribute. Therefore they do not have the incentive to bid to maximize total actions. Actually, a \emph{rational} DSP will always bid at the price
\begin{equation}
\text{eCPM}=\text{AR}\times \text{CPA} \times p(attribution|action)
\end{equation}
where $p(attribution|action)$ is the probability it gets attributed if an action happens. The industry common practice to bid AR$\times$CPA is simplifying this by assuming that the full credit of an action will be eventually attributed to the DSP. In many scenarios, such assumption is true. However, we must point out that such assumption is not always valid especially when multiple DSPs are running campaigns for the same advertiser simultaneously. \\
\indent Given that the DSPs will always bid a rational eCPM price, we are more interested in how advertisers can move them to lift-based bidding, which we have shown can bring more actions to the advertisers. Intuitively, if the DSPs are attributed based on the relative AR lift they contribute to the final AR, they have more incentive to practice lift-based bidding. Therefore the key is the attribution model. \\
\indent Again, let $u_i$ be the user of the $i$-th ad request, $p_i$ be the AR if the advertiser's ad is shown to $u_i$, and $\Delta p_i$ be the AR lift. Let $a_i=p(attribution|action, u_i)$ be the probability that the action from $u_i$ is attributed to the DSP that wins $u_i$. Suppose there are two DSPs $DSP_1$ and $DSP_2$ bidding on behalf of the same advertiser at the same time. $DSP_1$ always bid the rational price (i.e., $\text{CPA}\times p_i \times a_i$) while $DSP_2$ practices lift-based bidding (i.e., bid $\beta \times \Delta p_i$).

\begin{theorem}
Unless $a_i =  \frac{\beta}{CPA} \times \frac{\Delta p_i}{p_i}$ (i.e., $DSP_1$ always bid the same price as $DSP_2$), $DSP_2$ yields more actions than $DSP_1$ for the advertiser when the advertiser attributes the same amount of actions to them.
\begin{proof}
Let $i$ be the index of all the users, $j$ be the index of those users that $DSP_1$ wins (i.e., $\text{CPA} \times p_j \times a_j > \beta \times \Delta p_j$), and $k$ be the index of those users that $DSP_2$ wins (i.e., $\text{CPA}\times p_k \times a_k < \beta \times \Delta p_k$). It is straightforward to see that the expected number of actions to be attributed to $DSP_1$ and $DSP_2$ are $\sum_j p_j \times a_j$ and $\sum_k p_k \times a_k$ respectively. The expected number of actions if only $DSP_1$ is considered is $\sum_j p_j + \sum_k (p_k-\Delta p_k)$, and the expected number of actions if only $DSP_2$ is considered is $\sum_j (p_j - \Delta p_j) + \sum_k p_k$.\\
\indent Therefore, let $\mathscr{A}_1$ ($\mathscr{A}_2$) be the expected number of actions per attributed action if only $DSP_1$ ($DSP_2$) is considered, we have
\begin{equation}
\mathscr{A}_1=\frac{\sum_j p_j + \sum_k (p_k-\Delta p_k)}{\sum_j p_j \times a_j}
\end{equation}
\begin{equation}
\mathscr{A}_2=\frac{\sum_j (p_j - \Delta p_j) + \sum_k p_k}{\sum_k p_k \times a_k}
\end{equation}
\indent If $DSP_1$ and $DSP_2$ are not always bidding the same price, we can always adjust $\beta$ to control the winning landscape so that $DPS_1$ and $DSP_2$ get the same amount of attribution. When the same amount of actions is attributed to $DSP_1$ and $DSP_2$ (i.e., $\sum_j p_j \times a_j =\sum_k p_k \times a_k$), by swapping the denominators and consolidating the numerators in Equation 1 and 2, noticing $\text{CPA} \times p_j \times a_j > \beta \times \Delta p_j$ and $\text{CPA} \times p_k \times a_k < \beta \times \Delta p_k$, we have
\begin{equation}
\mathscr{A}_1=\frac{\sum_i p_i - \sum_k \Delta p_k}{\sum_k p_k \times a_k} < \frac{\sum_i p_i}{\sum_k p_k \times a_k} - \frac{\text{CPA}}{\beta}
\end{equation}
\begin{equation}
\mathscr{A}_2=\frac{\sum_i p_i - \sum_j \Delta p_j}{\sum_j p_j \times a_j} > \frac{\sum_i p_i}{\sum_j p_j \times a_j} - \frac{\text{CPA}}{\beta}
\end{equation}
\indent Therefore $\mathscr{A}_1 < \mathscr{A}_2$ unless $DSP_1$ always bid the same price as $DSP_2$ i.e., $a_i=\frac{\beta}{CPA} \times \frac{\Delta p_i}{p_i}$. 
\end{proof}
\end{theorem}
\begin{theorem}
Unless $a_i =  \frac{\beta}{CPA} \times \frac{\Delta p_i}{p_i}$ (i.e., $DSP_1$ always bid the same price as $DSP_2$), $DSP_2$ costs more than $DSP_1$ when the advertiser attributes the same amount of actions to them. 
\begin{proof}
The expected number of actions to be attributed to $DSP_1$ and $DSP_2$ are $\sum_j p_j \times a_j$ and $\sum_k p_k \times a_k$ respectively. The cost of $DSP_1$ and $DSP_2$ are $\sum_j \beta \times \Delta p_j$ and $\sum_k \text{CPA} \times p_k \times a_k$ respectively. Therefore let $\mathscr{C}_1$ and $\mathscr{C}_2$ be the cost per attributed action of $DSP_1$ and $DSP_2$ respectively,  we have
\begin{equation}
\mathscr{C}_1=\frac{\sum_j \beta \times \Delta p_j}{\sum_j p_j \times a_j}
\end{equation}
\begin{equation}
\mathscr{C}_2=\frac{\sum_k \text{CPA} \times p_k \times a_k}{\sum_k p_k \times a_k} 
\end{equation}
\indent Noticing $\text{CPA} \times p_j \times a_j > \beta \times \Delta p_j$, we have $\mathscr{C}_1 < \text{CPA}$. Since $\mathscr{C}_2=\text{CPA}$, $\mathscr{C}_1 < \mathscr{C}_2$ unless $DSP_1$ always place the same bid as $DSP_2$, i.e., $a_i =  \frac{\beta}{CPA} \times \frac{\Delta p_i}{p_i}$. 
\end{proof}
\end{theorem}
Theorem 3 and 4 suggest that for a rational DSP, the only way to move it to lift-based bidding is to attribute based on the relative lift it contributes to the final action (i.e., $a_i\propto \frac{\Delta p_i}{p_i}$). We believe by directly paying for the performance lift, the advertisers can move the DSPs to the more optimized bidding strategy.

\section{Related work}
Predicting AR for display ad has attracted much research interest. Most existing works towards this goal model actions and clicks within the same framework \cite{KLee:EstimateCVRTurn,rosales2012post,chapelle2014modeling}. We notice that none of these existing works explicitly models AR lift. Although researchers are exploring alternative bidding strategies in order to optimize some Key Performance Indicator (KPI) under budget constraint throughout the campaign's lifetime \cite{zhang2014optimal,perlich2012bid}, our work is fundamentally different from these bid optimization works. First, they assume a CPM pricing model and therefore bid optimization is under budget constraint, which is not a serious concern in CPA pricing model. Second, in their approaches, the modified bids are still derived from functions that map absolute AR to the final bid. So they still bid based on the absolute AR. Several data-driven multi-touch attribution methods have been proposed in recent years \cite{shao2011data,dalessandro2012causally,wooff2014time}. Our focus in this paper is more on illustrating the relationship between bidding strategies and attribution models than on a specific attribution method. Budget allocation based on multi-touch attribution is studied in \cite{geyik2014multi}, which we claim can be complementary to our work. After budget allocation is done, our approach can further optimize in the individual impression level. In \cite{dalessandro2012evaluating}, the authors also pointed out that some marketers may prefer to evaluate and optimize campaigns for incremental purchases. However, their focus was identifying better proxies than clicks to evaluate online advertising effectiveness instead of innovating bidding strategies. 

\section{Conclusion and Future Work}
We presented a general modeling framework to predict the AR lift, and apply it to drive a novel bidding strategy. We have proven both mathematically and empirically that the new bidding strategy indeed helps advertisers to improve campaign performance. We believe it will be an industry trend that advertisers will pay only for those impressions that drive incremental market values. This is an exciting new research area with many potential directions such as building accurate AR lift prediction models, innovating alternative CPA pricing models, and developing fairer attribution models. Another particularly interesting topic is how these models and technologies can be implemented in the industry where multiple parties such as advertisers, DSPs, ad exchanges, and third-party verification companies are all involved.

\bibliographystyle{aaai}
\bibliography{lift}

\end{document}